\documentclass[11pt]{article}
\usepackage[utf8]{inputenc}
\usepackage[english]{babel}
\usepackage{amsmath,amsfonts,amsthm,graphicx,float}
\usepackage{natbib}

\DeclareMathAlphabet{\mathpzc}{OT1}{pzc}{m}{it}

\newcommand{\tr} {\mbox{tr}}

\usepackage{algorithm}

\usepackage[noend]{algpseudocode}

\def\b1{{\mathbf 1}}

\usepackage[font=small]{caption}

\usepackage{titlesec}

\titleformat*{\section}{\normalfont\fontsize{14}{17}\bfseries}
\titleformat*{\subsection}{\normalfont\fontsize{12}{15}\bfseries}

\usepackage{bm}
\usepackage{enumerate}

\usepackage{soul}
\usepackage[usenames, dvipsnames]{color}

\newcommand{\verde}{\textcolor{Green}}

\date{}

\begin{document}

\title{\LARGE Nonparametric conditional risk mapping under heteroscedasticity}\normalsize
\author{Rub\'en Fern\'andez-Casal \\
Universidade da Coru\~{n}a\thanks{%
Research group MODES, CITIC, Department of Mathematics, Faculty of Computer Science, Universidade da Coru\~na, Campus de Elvi\~na s/n, 15071,
A Coru\~na, Spain}
\and
Sergio Castillo-P\'aez \\
Universidad de las Fuerzas Armadas ESPE\thanks{Departamento de Ciencias Exactas, Universidad de las Fuerzas Armadas ESPE, Av. General Rumi\~nahui s/n, 171103, Sangolqu\'i, Ecuador}
\and %
Mario Francisco-Fern\'andez\\
Universidade da Coru\~{n}a\footnotemark[1]
}
\maketitle


\begin{abstract}
A nonparametric procedure to estimate the conditional probability that a nonstationary geostatistical process exceeds a certain threshold value is proposed. The method consists of a bootstrap algorithm that combines conditional simulation techniques with nonparametric estimations of the trend and the variability. The nonparametric local linear estimator, considering a bandwidth matrix selected by a method that takes the spatial dependence into account, is used to estimate the trend. The variability is modeled estimating the conditional variance and the variogram from corrected residuals to avoid the biasses. The proposed method allows to obtain estimates of the conditional exceedance risk in non-observed spatial locations. The performance of the approach is analyzed by simulation and illustrated with the application to a real data set of precipitations in the U.S.

\end{abstract}	


\section{Introduction}
\label{sec_intro}

Risk maps containing the probabilities that a certain variable of interest exceeds a given threshold or permissible value in an area of study are usually employed by environmental agencies to control different pollution levels (in soil, air or water) or to alert population of possible natural disasters (earthquakes, floods, etc.). The estimation of these exceeding probabilities using simple and reliable statistical methods is, therefore, an important practical issue. The resulting estimated maps can help governments to make decisions and to organize prevention policies in the near future. 

\begin{figure}[h]
    \centering
		\includegraphics[width=0.75\textwidth]{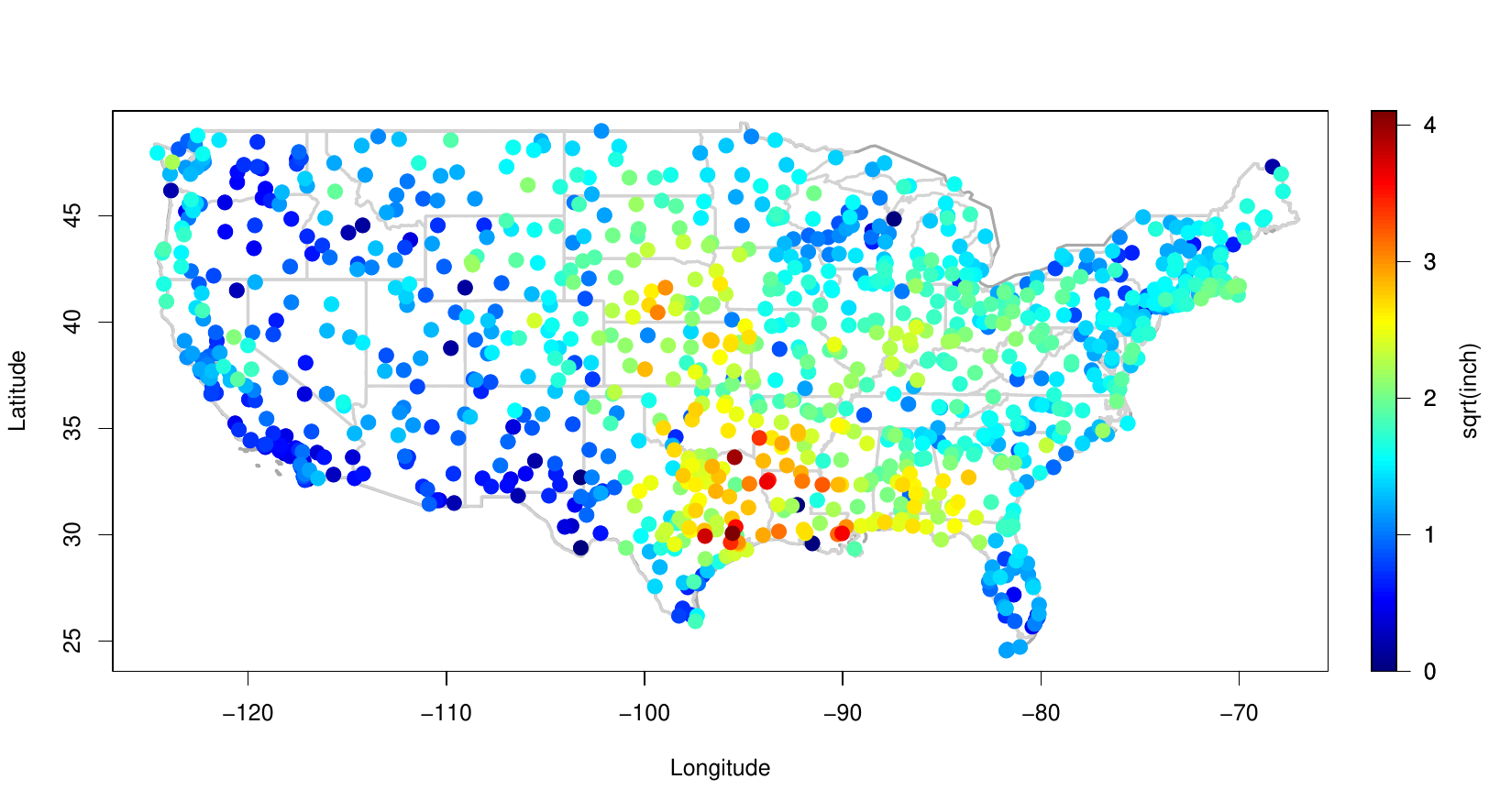}
	 \caption{Spatial locations and observed values of the total precipitations
	 	(square-root of rainfall inches) during March 2016 recorded over
	 	1053 locations on the continental part of USA.}
    \label{fig_obser}
\end{figure}

In this paper, we develop a nonparametric methodology to produce risk maps and apply it to 
a data set containing the total precipitations (square-root of rainfall inches) during March 2016 recorded over 1053 locations on the continental part of the U.S. (Figure \ref{fig_obser} contains the observed values). The goal is to estimate the conditional probability of occurring a total precipitation larger than or equal to a threshold value, which could have a direct application in agriculture or in flood prevention, for example.

Different geostatistical techniques have usually been employed to approximate exceeding probabilities. These methods include traditional approaches, such as the indicator kriging (IK) \citep[e.g.][]{G97} or the disjunctive kriging (DK) \citep[e.g.][]{W89}, or more recent procedures, such us those based on analysis of compositional data \citep[e.g.][]{T08}. 
The IK consists in the application of the ordinary kriging linear predictor to indicator functions of the data. Although it is perhaps  the most popular approach in this context, it has some drawbacks. {First}, the discretization of the data can lead to a loss of information. On the other hand, the estimated probabilities could be greater than one or negative. Moreover, it could present order-relation problems \citep[see, e.g.][Section 6.3.3]{C12}. Some of these issues can be avoided with the use of the so-called simplicial indicator kriging \citep{T08}. This method employs a simplex approach for compositional data to estimate  the conditional cumulative distribution function. Another alternative to the IK is the DK, a nonlinear estimation technique which usually assumes a Gaussian isofactorial model for the geostatistical process. However, there is no empirical evidence to recommend the DK in preference to the IK, or the opposite \citep{Lark2004}. 



The approaches previously described usually suppose stationarity and a parametric {model. Therefore, if the assumed mo\-del is not appropriate, the conclusions drawn may be unreliable or even wrong}. To avoid these problems, alternatively, nonparametric techniques could be used. 
For instance, \cite{Pilar17} proposed two kernel-based estimators to approximate the local distribution under homoscedasticity. 
In this line, \cite{FC18} proposed an unconditional bootstrap method to estimate the spatial risk, without assuming any parametric form for the trend function nor for the dependence structure of the process. They consider a homoscedastic model and used local linear estimates of the trend and the variogram, jointly with a procedure to correct the bias introduced by the direct use of the
residuals in the variogram estimation.
On the other hand, this nonparametric procedure was extended to a heteroscedastic context in \cite{FC20}. In both cases, although the use of nonparametric methods avoids misspecification problems, they focus on the estimation of the unconditional probability that the variable under study exceeds a threshold. 
{Note that, as the replicas obtained by unconditional simulation will not necessarily match the observed sample values \citep[see, e.g.][Chapter 7]{C12}, a direct comparison of these procedures with the ones described in the previous paragraph is not entirely appropriate, since they aim at the estimation of conditional exceeding probabilities}. 

In the present work, we propose a bootstrap method to estimate threshold exceeding conditional probabilities
under heteroscedasticity of the spatial process. This approach uses the unconditional bootstrap method introduced in \cite{FC20} as part of its implementation. The new procedure generates condicional replicates matching up the observed values at the sampled locations. The conditionalization of simulations {is equivalent to choose} among all possible unconditional simulations of the spatial process, those that coincide with the values obtained at the observation locations \citep{J74}.

The remainder of the paper is organized as follows. In Section \ref{sec_np_modeling}, the spatial model considered in this research is presented. Additionally, the nonparametric estimators for the mean or trend function, the variance and the variogram employed in the conditional bootstrap method are introduced. 
In Section \ref{sec_boot} the bootstrap algorithm to estimate the conditional risk is described (specifically, in Section  \ref{sec_cond_boot}).
In this procedure, a slight modification of the bootstrap method proposed in \cite{FC20} to approximate the unconditional risk (also discussed in Section \ref{sec_uncond_boot}) is used.
A simulation study for assessing the performance of the new approach, considering stationary and nonstationary processes, under regular and non-regular sampling designs, is provided in Section \ref{sec_simulation}. 
Section \ref{sec_app_data} discusses the application of the methods to the precipitation data introduced above. 
Finally, Section \ref{sec_conclus} contains  some conclusions and finals remarks.

\section{Nonparametric modeling}
\label{sec_np_modeling}

Suppose that $\left\{ Y(\mathbf{x}),\mathbf{x}\in D \subset \mathbb{R}^{d}\right\}$ is a spatial heteroscedastic process which can be modeled as follows:
\begin{equation}  \label{hetermodel}
	Y(\mathbf{x})=\mu(\mathbf{x})+\sigma(\mathbf{x})\varepsilon(\mathbf{x}),
\end{equation} 
where $\mu(\cdot)$ and $\sigma^2(\cdot)$ are the trend and variance functions, and $\varepsilon(\cdot)$ is a second order stationary process with zero mean, unit variance and correlogram  $\rho(\mathbf{u}) = {\rm Cov}\left[ \varepsilon ( \mathbf{x}), \allowbreak \varepsilon (\mathbf{x}+\mathbf{u}) \right] $, for $\mathbf{x}$ and $\mathbf{x} + \mathbf{u} \in D$.
No specific forms will be assumed for $\mu(\cdot)$, $\sigma^2(\cdot)$ and $\rho(\cdot)$, although they should be smooth functions to be consistently estimated.

The goal is estimate nonparametrically the conditional probability
\begin{equation}
	r_c(\mathbf{x}_{\alpha}^e, \mathbf{Y})=P\left[Y(\mathbf{x}_{\alpha}^e)\geq c \mid \mathbf{Y}\right],
	\label{cond_prob}
\end{equation}
where $\mathbf{Y}=\left[ Y(\mathbf{x}_{1}),\ldots,Y(\mathbf{x}_{n}) \right]^{t}$ are observed values of the process at certain sample locations, $c$ is a threshold (critical) value, and $\left\lbrace \mathbf{x}_{\alpha}^e \right\rbrace_{\alpha=1}^{n_0}$ is a set of unobserved estimation locations.

 It must be taken into account that the spatial dependence of the process $Y$ depends on the variance and the correlogram of $\varepsilon$.
For instance, the covariance matrix of the observations $\mathbf{Y}$ can be expressed as:
\begin{equation*}
\boldsymbol{\Sigma} =  \mathbf{DRD},	
\end{equation*}
where $\mathbf{D}= {\rm diag}\left[\sigma(\mathbf{x}_{1}),\ldots ,\sigma(\mathbf{x}_{n})\right]$ and $\mathbf{R}$ is the correlation matrix of the (unknown) errors $\boldsymbol{\varepsilon} = \left[\varepsilon(\mathbf{x}_{1}), \ldots ,\varepsilon(\mathbf{x}_{n})\right]^t$. The latter matrix is usually estimated from the semivariogram $\gamma(\mathbf{u})= \tfrac{1}{2}Var\left[\varepsilon \left( \mathbf{x}\right)-\varepsilon \left(\mathbf{x}+\mathbf{ u}\right) \right] = 1 - \rho(\mathbf{u})$

The first step in the proposed approach consists in the nonparametric estimation of the trend, the variance and the dependence of the spatial process.
Different nonparametric approaches have been used for the estimation of the model components, including kernel methods, splines or wavelets techniques. For instance, a comprehensive revision of trend estimation approaches can be found in \cite{O01}. In that paper, the authors focus on the framework of regression models with correlated homoscedastic errors.
Nonparametric methods for the estimation of the functional variance have mainly based on the approximation of the mean of the squared residuals \citep[see e.g.][for the independent case]{Fan98}. 
In this line and among the available literature, we may highlight the works of \cite{ruppert1997}, who proposed a degrees-of-freedom correction of the bias due to the preliminary estimation of the trend with uncorrelated errors, and \cite{Vilar2006}, who studied the properties of the squared residual estimator for one-dimensional correlated data.
In the present paper, assuming model (\ref{hetermodel}), a similar procedure to that proposed in \cite{FC17} is used to nonparametrically estimate the model components.

To estimate the trend $\mu(\cdot)$, we consider a kernel-based method called the local linear estimator. This approach has shown a very good performance from a theoretical and a practical point of view \citep[e.g.][]{Fan1996}.
In the spatial framework, given a sample $\{[\mathbf{x}_{i},Y(\mathbf{x}_{i})]\}_{i=1}^{n}$, the local linear estimator of $\mu(\mathbf{x})$ is given by:
\begin{equation}
\label{trend_est}   
    \hat{\mu}_{\mathbf{H}}(\mathbf{x})= \mathbf{e}_{1}^{t} \left( \mathbf{X}_{\mathbf{x}}^{t}{\mathbf{W}}_{\mathbf{x}}\mathbf{X}_{\mathbf{x}}
    \right)^{-1}
    \mathbf{X}_{\mathbf{x}}^{t}{\mathbf{W}}_{\mathbf{x}} \mathbf{Y}
    \equiv {s}_{\mathbf{x},\mathbf{H} }^{t}\mathbf{Y},
\end{equation}
where $\mathbf{e}_{1}$ is a vector with $1$ in the first entry and all other entries 0, $\mathbf{X}_{\mathbf{x}}$ is a matrix 
whose $i$-th row is $[1,(\mathbf{x}_{i}-\mathbf{x})^{t}]$, 
$\mathbf{W}_{\mathbf{x}}=\mathtt{diag}\left[ K_{\mathbf{H}}(\mathbf{x}_{1}-\mathbf{x}),\ldots,K_{\mathbf{H}}(\mathbf{x}_{n}-\mathbf{x})\right]$, with $K_{\mathbf{H}}(\mathbf{u})=\left\vert \mathbf{H} \right\vert ^{-1}K(\mathbf{H}^{-1}\mathbf{u})$, being $K$ a multivariate kernel function, and $\mathbf{H}$ is a $d\times d$ nonsingular symmetric matrix called the bandwidth matrix. Note that $\hat{\mu}_{\mathbf{H}}(\mathbf{x})$ is a linear smoother, since the estimated values at the sample locations can be expressed as $\hat{\boldsymbol{\mu}}= \left[\hat{\mu}_{\mathbf{H}}(\mathbf{x}_1), \ldots, \hat{\mu}_{\mathbf{H}}(\mathbf{x}_n)\right]^t = \mathbf{S}_{\mathbf{H}}\mathbf{Y}$, 
being $\mathbf{S}_{\mathbf{H}}$ the smoothing matrix whose $i$-th row is equal to ${s}_{\mathbf{x}_i,{\mathbf{H}}}^{t}$ (the smoother vector for $\mathbf{x} = \mathbf{x}_{i}$).

On the other hand, the usual nonparametric procedure for estimation of the small-scale structure of the process is carried out from the residuals $\mathbf{r}=\left(r_{1},\ldots,r_{n}\right)^t=\mathbf{Y}-\mathbf{S}_{\mathbf{H}}\mathbf{Y}$. A variance estimate is obtained by linear smoothing the squared residuals $\{[\mathbf{x}_{i},r^2_{i}]\}_{i=1}^{n}$, 
\begin{equation}
\label{var_pilot}
\hat{{\sigma}}^2_{\mathbf{r},\mathbf{H}_2}(\mathbf{x}) = {s}_{\mathbf{x},\mathbf{H}_2 }^{t}\mathbf{r}^2,
\end{equation}
where $\mathbf{r}^2=\left( r^2_{1},\ldots,r^2_{n} \right)^t$ and $\mathbf{H}_2$ is the corresponding bandwidth matrix. Likewise, a (pilot) residual semivariogram estimate $\hat{\gamma}_{\hat{\boldsymbol{ \varepsilon}}}(\cdot)$ is obtained by linear smoothing the sample semivariances
\begin{equation}
\label{variog_linloc}
\left\{\left(||\mathbf{x}_{i}-\mathbf{x}_{j }||,\left[\hat{\varepsilon}(\mathbf{x}_{i})- \hat{\varepsilon}(\mathbf{x}_{j})\right]^2 \right) :1\le i < j \le n\right\}
\end{equation}
of the (estimated) standardized residuals $\hat{\boldsymbol{\varepsilon}} = [\hat{\varepsilon}(\mathbf{x}_{1}), \ldots, \hat{\varepsilon}(\mathbf{x}_{n})]^t =  \hat{\mathbf{D}}^{-1}_0 \mathbf{r}$, being $$\hat{\mathbf{D}}_0 = \allowbreak {\rm diag}\left[ \allowbreak \hat{\sigma}_{\mathbf{r}, \mathbf{H}_2}(\mathbf{x}_{1}), \ldots, \hat{\sigma}_{\mathbf{r},\mathbf{H}_2}(\mathbf{x}_{n})\right].$$
In this case, assuming isotropy for simplicity, it would only be necessary to consider a scalar bandwidth parameter $h_3$.

Bandwidth parameters play an important role in the performance of the \verde{previous} local linear estimators, since they control the shape and the size of the local neighborhoods used to obtain the corresponding estimates, determining their smoothness. When the data are spatially correlated, as it is assumed in the present paper, we recommend the use of the ``bias corrected and estimated'' generalized cross-validation (CGCV) criterion, proposed in \cite{F05}, to select the matrices $\mathbf{H}$ and $\mathbf{H}_2$, by minimizing 
\begin{equation*}
{\rm CGCV}(\mathbf{H}) = \frac{1}{n} \sum_{i=1}^{n}\left[ \frac{Y(\mathbf{x}_{i}) - \hat{\mu}_{\mathbf{H}}(\mathbf{x}_{i})}{1-\frac{1}{n}\tr\left(\mathbf{S}_\mathbf{H}\hat{\mathbf{R}}\right)}\right]^{2}
\label{gcvce}
\end{equation*}
and
\begin{equation*}
{\rm CGCV}(\mathbf{H}_2) = \frac{1}{n} \sum_{i=1}^{n}\left[ 
\frac{r_i^2 - \hat{\sigma}^2_{\mathbf{r},\mathbf{H}_2}(\mathbf{x}_{i})} {1-\frac{1}{n}\tr\left(\mathbf{S}_{\mathbf{H}_2} \hat{\mathbf{R}}_{\mathbf{r}^2}\right)}\right]^{2}, 
\end{equation*}
respectively, where $\tr(\mathbf{A})$ stands for the trace of a square matrix $\mathbf{A}$, and $\hat{\mathbf{R}}$ and $\hat{\mathbf{R}}_{\mathbf{r}^2}$ are estimates of the correlation matrices of the observations and of the squared residuals, respectively. 
A simpler approximation of the covariance matrix of the squared residuals $\boldsymbol{\Sigma}_{\mathbf{r}^2}$ can be obtained under the assumptions of normality and zero mean for the residuals. In that case,  
$$\boldsymbol{\Sigma}_{\mathbf{r}^2} = 2\boldsymbol{\Sigma}_{\mathbf{r}}\odot\boldsymbol{\Sigma}_{\mathbf{r}},$$
where $\odot$ represents the Hadamard product and $\boldsymbol{\Sigma}_{\mathbf{r}} = Var(\mathbf{r})$ \citep{ruppert1997}.
Finally, the bandwidth parameter $h_3$ for the computation of the residual semivariogram estimate $\hat{\gamma}_{\hat{\boldsymbol{ \varepsilon}}}(\cdot)$ can be selected as the minimizer of the cross-validation relative squared error
$$
\sum\limits_{i=1}^{n-1} \sum\limits_{j=i+1}^{n} \left[ \frac{ \left(\hat{\varepsilon} (\mathbf{x}_{i})-\hat{\varepsilon}(\mathbf{x}_{j}) \right)^2 }{2 \hat{\gamma}^{-(i,j)}_{\hat{\boldsymbol{ \varepsilon}}}\left(||\mathbf{x}_{i}-\mathbf{x}_{j} ||\right) } - 1\right]^2,
$$
where $\hat{\gamma}^{-(i,j)}_{\hat{\boldsymbol{ \varepsilon}}}$ is the estimate obtained when excluding the pair $(i,j)$ in (\ref{variog_linloc}).

Nevertheless, it is well known that estimates based on direct use of residuals underestimate the variability of the spatial process:
\[\boldsymbol{\Sigma}_{\mathbf{r}} = \boldsymbol{\Sigma} +
\mathbf{S}_\mathbf{H} \boldsymbol{\Sigma} \mathbf{S}_\mathbf{H}^{t} - \boldsymbol{\Sigma} \mathbf{S}_\mathbf{H}^{t} - \mathbf{S}_\mathbf{H} \boldsymbol{\Sigma}\]
\citep[see, e.g.][Section 3.4.3, for the linear case under homoscedasticity]{C93}.
Equivalently,
$$\begin{array}{l}
{\mbox{Var}}\left(r_i\right) = \sigma^2(\mathbf{x}_{i})\left(1 + b_{ii}\right),\\
{\mbox{Var}}\left[r_i/\sigma(\mathbf{x}_{i}) - r_j/\sigma(\mathbf{x}_{j}) \right] =
{\mbox{Var}}\left[\varepsilon(\mathbf{x}_i) - \varepsilon(\mathbf{x}_j) \right]+ b_{ii} + b_{jj} - 2 b_{ij},
\end{array}$$
where $b_{ij}$ is the $(i,j)$-th element of $$\mathbf{B} = \mathbf{D}^{-1} \left(\mathbf{S}_{\mathbf{H}} \boldsymbol{\Sigma} \mathbf{S}_{\mathbf{H}}^{t} - \boldsymbol{\Sigma} \mathbf{S}_{\mathbf{H}}^{t} - \mathbf{S}_{\mathbf{H}} \boldsymbol{\Sigma} \right)  \mathbf{D}^{-1},$$ a square matrix representing the bias due to the direct use of the residuals. 
As it may have a significant impact on risk assessment, a slight modification of the iterative procedure proposed in \cite{FC20} is used to obtain approximately unbiased estimates of the variance $\sigma^2(\cdot)$ and the error variogram $\gamma(\cdot)$.
Starting with the residual estimates, $\hat{{\sigma}}^2_{\mathbf{r},\mathbf{H}_2}(\mathbf{x})$ and $\hat{\gamma}_{\hat{\boldsymbol{ \varepsilon}}}(\cdot)$. At each iteration, the bias matrix $\mathbf{B}$ is approximated by $\hat{\mathbf{B}} =\hat{\mathbf{D}}^{-1} (\mathbf{S}_{\mathbf{H}} \hat{\boldsymbol{\Sigma}} \mathbf{S}_{\mathbf{H}}^{t} - \hat{\boldsymbol{\Sigma}} \mathbf{S}_{\mathbf{H}}^{t} - \mathbf{S}_{\mathbf{H}} \hat{\boldsymbol{\Sigma}} )  \hat{\mathbf{D}}^{-1}$. Then, an updated estimate $\hat{{\sigma}}^2(\cdot)$ is computed by replacing $r_i^2$ by $r_i^2/(1+\hat{b}_{ii})$ in (\ref{var_pilot}), and a ``corrected" $\hat{\gamma}(\cdot)$ is derived by substituting $\left[\hat{\varepsilon}(\mathbf{x}_{i})- \hat{\varepsilon}(\mathbf{x}_{j})\right]^2$ for $\left[\hat{\varepsilon}(\mathbf{x}_{i})- \hat{\varepsilon}(\mathbf{x}_{j})\right]^2-\hat{b}_{ii}-\hat{b}_{jj}+2\hat{b}_{ij}$ in (\ref{variog_linloc}).

Note that the pilot local linear variogram estimates, $\hat{\gamma}_{\hat{\boldsymbol{ \varepsilon}}}(\cdot)$ and $\hat{\gamma}(\cdot)$, obtained with the above procedure are not necessarily conditionally negative definite functions and cannot be directly used for prediction or simulation. Valid variogram estimates are obtained by fitting ``nonparametric'' isotropic Shapiro-Botha models \citep{SB91} to the pilot estimates \citep[see e.g.][Section 4, for a description of this algorithm]{FC17}, which will be denoted by $\bar{\gamma}_{\hat{\boldsymbol{ \varepsilon}}}(\cdot)$ and $\bar{\gamma}(\cdot)$, respectively.

\section{Unconditional and Conditional bootstrap algorithms}
\label{sec_boot}

In this section, the bootstrap algorithm to estimate  the conditional risk under heteroscedasticity is presented. This method is based on a general conditional simulation method combining unconditional simulations with kriging predictions \citep[see, e.g.][Section 7.3.1]{C12}. In a first step, the bootstrap algorithm studied in \cite{FC20}, and described below, is used to generate the unconditional replicas. 

\subsection{Unconditional bootstrap algorithm}
\label{sec_uncond_boot}

The present bootstrap algorithm is used to generate unconditional replicates $Y^*_{NS}(\mathbf{x}_{\alpha}^e)$ at the different estimation locations $\left\lbrace \mathbf{x}_\alpha^e: \alpha=1,\ldots,n_0 \right\rbrace$, following these steps:

\begin{enumerate}
	
	\item Using the procedure described in the previous section:
	
	\begin{enumerate}
		\item Obtain $\hat{\mu}_{\mathbf{H}}(\cdot)$, the corresponding residuals $\mathbf{r}$, the initial $\hat{{\sigma}}^2_{\mathbf{r},\mathbf{H}_2}(\mathbf{x})$ and final $\hat{{\sigma}}^2(\cdot)$ variance estimates, as well as the initial $\bar{\gamma}_{\hat{\boldsymbol{ \varepsilon}}}(\cdot)$ and final $\bar{\gamma}(\cdot)$ semivariogram estimates.
		
		\item Construct the matrix $\hat{\mathbf{R}}_0$
		from the residual variogram $\bar{\gamma}_{\hat{\boldsymbol{ \varepsilon}}}(\cdot)$, and obtain the Cholesky decomposition $\hat{\mathbf{R}}_0= \mathbf{L}_0\mathbf{L}_0^{t}$.
		
		\item Compute $\hat{\mathbf{R}}_\alpha$ corresponding to $\mathbf{x}_\alpha^e$ using $\bar{\gamma}(\cdot)$, and  $\mathbf{L}_\alpha$ such that $\hat{\mathbf{R}}_\alpha = \mathbf{ L}_\alpha\mathbf{L}_\alpha^t$.
		
		\item Construct the ``uncorrelated'' errors
		$\mathbf{e}= \mathbf{L}_{0}^{-1}\hat{\mathbf{D}}_0^{-1} \mathbf{r}$ and standardize them.
		
	\end{enumerate}

	\item Generate the unconditional bootstrap replicas as follows:
	
	\begin{enumerate}			
		
		\item Obtain independent bootstrap residuals of size $n_0$ from $\mathbf{e}$, denoted by $\mathbf{e}^{\ast}.$
		
		\item Compute the unconditional bootstrap residuals $\boldsymbol{\varepsilon}^*_{NC}={\mathbf{L}_\alpha}\mathbf{e}^{\ast }$.
		
		\item Construct the unconditional bootstrap replicas $$Y^*_{NC}(\mathbf{x}_{\alpha}^e)=\hat{\mu}_{\mathbf{H}}(\mathbf{x}_{\alpha}^e)
        + \hat{\sigma}(\mathbf{x}_{\alpha}^e)
        \boldsymbol{\varepsilon}^*_{NC}(\mathbf{x}_{\alpha}^e), \ \alpha=1, \ldots , n_0, $$
        being  $\boldsymbol{\varepsilon}^*_{NC}(\mathbf{x}_{\alpha}^e)$ the $\alpha$-th component of the vector ${\boldsymbol{\varepsilon}}^*_{NC}$.
	\end{enumerate}

\end{enumerate}


The previous algorithm produces bootstrap replicas that have their mean and variance-covariance matrix equal to the corresponding estimates of the spatial process $Y(\cdot)$ \citep[see, e.g.][Section 3.6.1]{C93}. 
However, as these replicas mimic an unconditional realization of the process and their values at the observation positions are random, they will not necessarily match the observed values  $\mathbf{Y}$ at the sample positions \citep[for more details, see e.g.][Chapter 7 ]{C12}.
Therefore, this algorithm is appropriate for estimating the unconditional risk, $P\left[Y(\mathbf{x}_{\alpha}^e)\geq c \right]$, but should not be used for the estimation of the conditional risk (\ref{cond_prob}), unless it is modified properly, for example, as shown below.

\subsection{Conditional bootstrap algorithm}
\label{sec_cond_boot}

Next, the proposed bootstrap algorithm to estimate the conditional risk \eqref{cond_prob} is described. The procedure uses unconditional replicas generated with the previous algorithm, although  it would not be necessary to obtain replicas of the whole process (Step 2-c above), only of the heteroscedastic errors
$$\delta^*_{NC}(\mathbf{x}_{\alpha}^e) = \hat\sigma(\mathbf{x}_{\alpha}^e){\varepsilon}^*_{NC}(\mathbf{x}_{\alpha}^e).$$

First of all, we will describe the principle of conditional simulation from a theoretical point of view.
For this, we would have to assume that the components of model (\ref{hetermodel}) (the trend, the variance and the variogram) are known, and the true errors $\boldsymbol{\varepsilon}=[\mathbf{\varepsilon} (\mathbf{x}_1),\ldots,\mathbf{\varepsilon}(\mathbf{x}_n)]^t$ are observed.
Obviously, these assumptions are unrealistic in a practical situation.
In fact, as described below, the theoretical components will be replaced by their estimates when using these ideas in the bootstrap method.
The conditional simulation of the error at a location $\mathbf{x}_{\alpha}^e$ \citep[see, e.g.][]{J74} is based on the trivial decomposition
\begin{equation}
\delta(\mathbf{x}_{\alpha}^e) = \hat{\delta}(\mathbf{x}_{\alpha}^e) + \left[\delta(\mathbf{x}_{\alpha}^e) - \hat{\delta}(\mathbf{x}_{\alpha}^e) \right], 
\label{aprox_sim}
\end{equation}
where $\hat{\delta}(\mathbf{x}_{\alpha}^e)$ is the simple kriging prediction at $\mathbf{x}_{\alpha}^e$ computed from $\boldsymbol{\delta} = [\sigma(\mathbf{x}_1){\varepsilon}(\mathbf{x}_1), \ldots, \allowbreak \sigma(\mathbf{x}_n){\varepsilon}(\mathbf{x}_n)]^t $. The idea is to substitute the unknown kriging error (the second term on the right hand side of \eqref{aprox_sim}) by a simulation of this error, obtained from an unconditional simulation $\delta_{NC}(\mathbf{x})$ of the error process. Then, a conditional simulation of this error is:
\begin{equation}
\delta_{CS}(\mathbf{x}_{\alpha}^e) = \hat{\delta}(\mathbf{x}_{\alpha}^e) + \left[\delta_{NC}(\mathbf{x}_{\alpha}^e) - \hat{\delta}_{NC}(\mathbf{x}_{\alpha}^e) \right],
\label{aprox_sim_cond}
\end{equation}  
where $\hat{\delta}_{NC}(\mathbf{x}_{\alpha}^e)$ is the kriging prediction at $\mathbf{x}_{\alpha}^e$ obtained from the unconditional simulations $\delta_{NC}(\mathbf{x}_i)$, $i=1, \ldots, n$, at the sample locations.
Proceeding in this way, it is easy to verify that $\delta_ {CS}(\mathbf{x}_i) = \mathbf{\delta}(\mathbf{x}_i)$ and, in the case of simple kriging, ${\mbox Var}\left[ \delta_ {CS}(\mathbf{x}) \right] = \sigma^2(\mathbf{x})$ and $$\mbox{Corr}[\delta_ {CS}(\mathbf{x}), \delta_ {CS}(\mathbf{x + u})] = \rho(\mathbf {u})$$ \citep[see, e.g.][Section 7.3.1]{C12}. These properties guarantee that the simulations reproduce the second order structure of the spatial process (and the complete distribution if, for instance, Gaussian errors are assumed). Note also that the simple kriging predictor is not being used because of its properties as an optimal linear predictor. It is simply a tool to incorporate the conditional covariances, assuming that they are known, in the matrix computations.

In the bootstrap world, the estimates of the model components play the role of the theoretical ones (the trend, the variance, the variogram and the true errors are known) and the previous results can be applied. 
Taking this into account, the proposed bootstrap algorithm to estimate the conditional risk is as follows:
	\begin{enumerate}
	
	\item Generate the unconditional bootstrap replicates at the estimation locations ${\delta}^*_{NC}(\mathbf{x}_\alpha^e)$, $\alpha=1,\ldots,n_0$, as well as in the sample locations ${\delta}^*_{NC}(\mathbf{x}_i)$, $i=1,\ldots,n$.
	
	\item Using simple kriging, obtain the predictions $\hat{\delta}(\mathbf{x}_\alpha^e)$ and $\hat{\delta}^*_{NC}(\mathbf{x}_\alpha^e)$ from the observed residuals $r(\mathbf{x}_{i})$ and the unconditional heteroscedastic errors ${\delta}^*_{NC}(\mathbf{x}_i)$, respectively.
	
	\item Calculate the conditional bootstrap heteroscedastic errors $${\delta}^*_{CS}(\mathbf{x}_\alpha^e) = \hat{\delta}(\mathbf{x}_\alpha^e) + \left[{\delta}^*_{NC}(\mathbf{x}_\alpha^e) - \hat{\delta}^*_{NC}(\mathbf{x}_\alpha^e) \right]. $$
	
	\item Construct the conditional bootstrap replicates $Y^*_{CS}(\mathbf{x}_\alpha^e)= \hat{\mu}_{\mathbf{H}}(\mathbf{x}_\alpha^e)+ {\delta}^*_{CS}(\mathbf{x}_\alpha^e)$.
	
	\item Repeat steps 1 to 4 a large number $B$ of times, to get $Y^{*(1)}_{CS}(\mathbf{x}_{\alpha}^e), \allowbreak \ldots, \allowbreak Y^{*(B)}_{CS}(\mathbf{x}_{\alpha}^e)$.
	
	\item Finally, estimate the conditional probability \eqref{cond_prob} by:
    \begin{equation}
    \label{cond_prob_est}
    \hat{r}_c(\mathbf{x}_\alpha^e, \mathbf{Y})=\frac{1}{B}\sum\limits_{j=1}^{B} \mathbb{I} \left[  {Y}^{*(j)}_{CS}(\mathbf{x}_{\alpha}^e)\geq c \right] ,
    \end{equation}
    where $\mathbb{I}(\cdot)$ represents the indicator function.

\end{enumerate}

\section{Simulation studies}
\label{sec_simulation}

\begin{figure}[h]
    \centering
	\begin{tabular}{cc}
		\footnotesize{(a)} & \footnotesize{(b)} \\
		\includegraphics[scale=0.27]{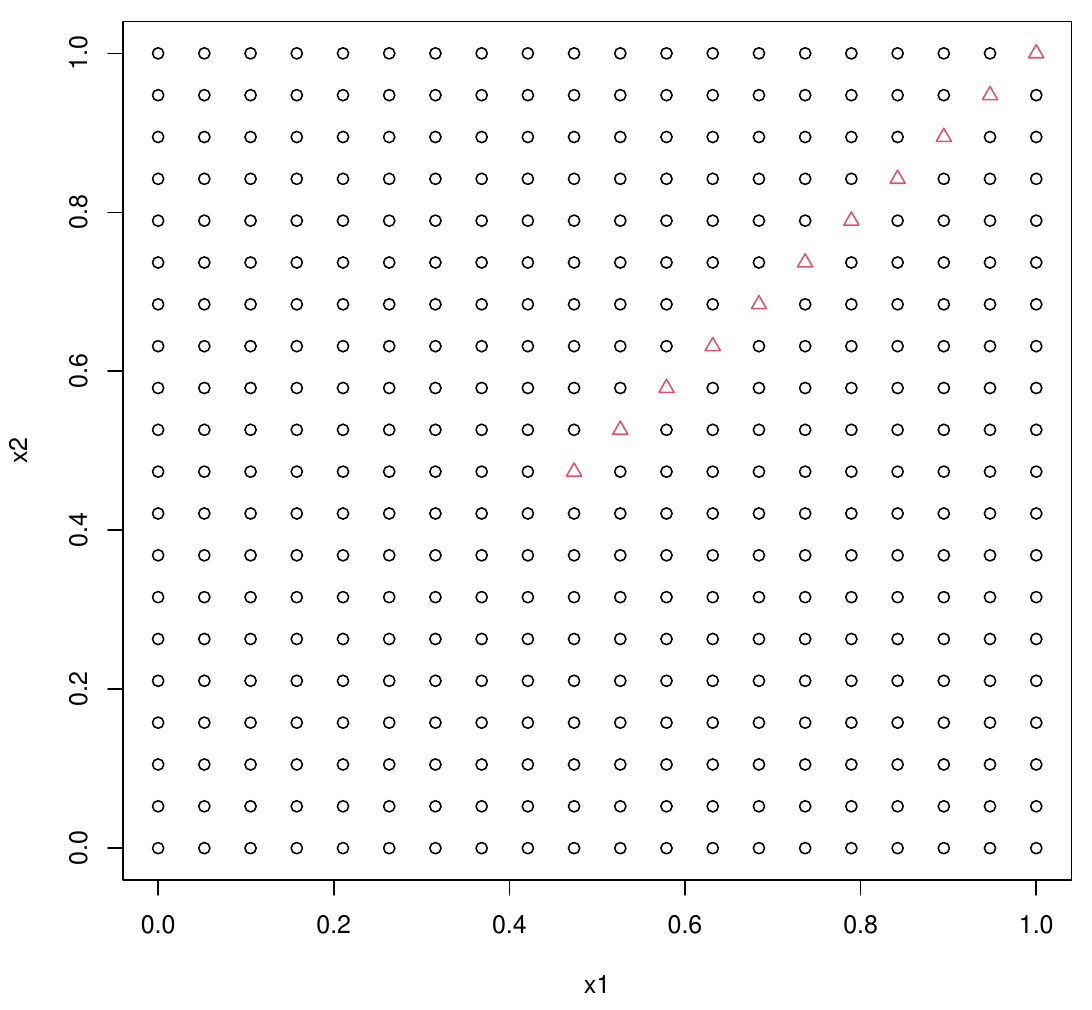}&
		\includegraphics[scale=0.3]{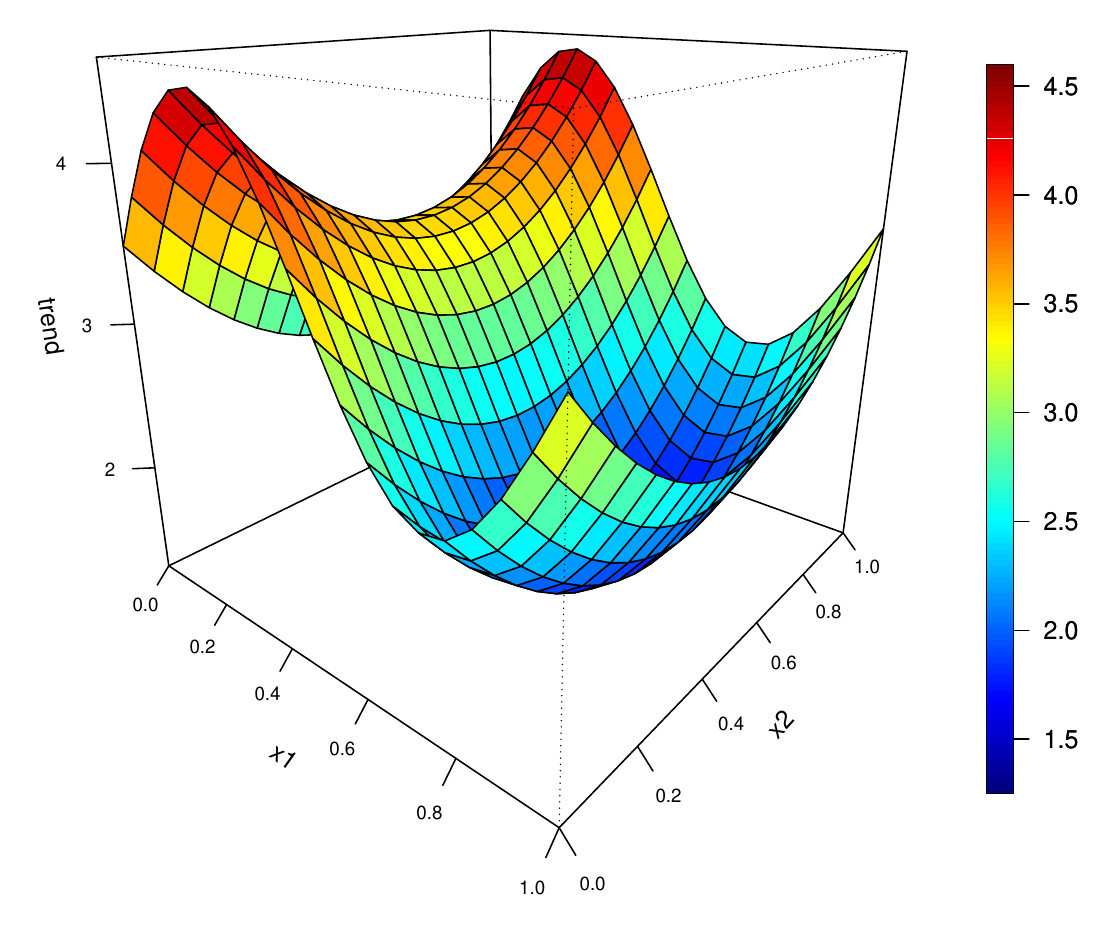}\\ 
		\footnotesize{(c)} & \footnotesize{(d)} \\
		\includegraphics[scale=0.3]{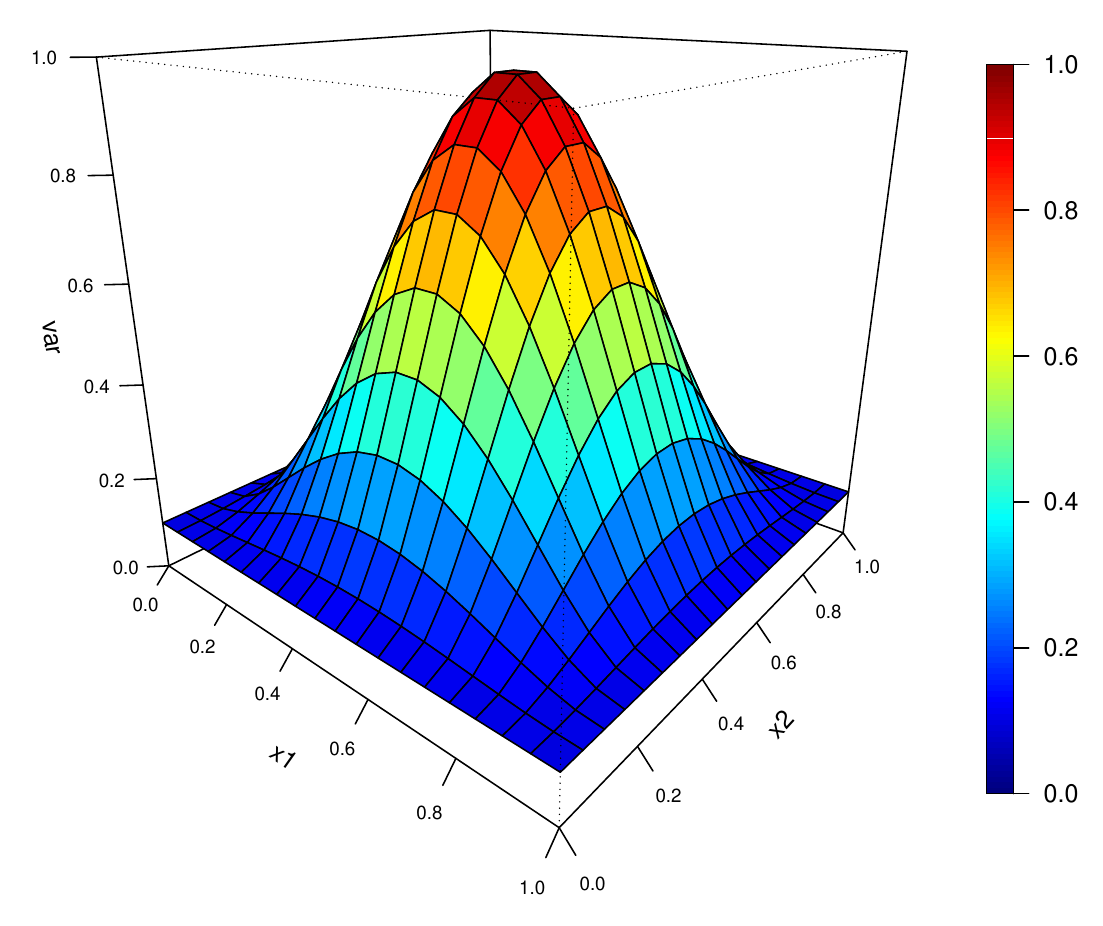}&
		\includegraphics[scale=0.3]{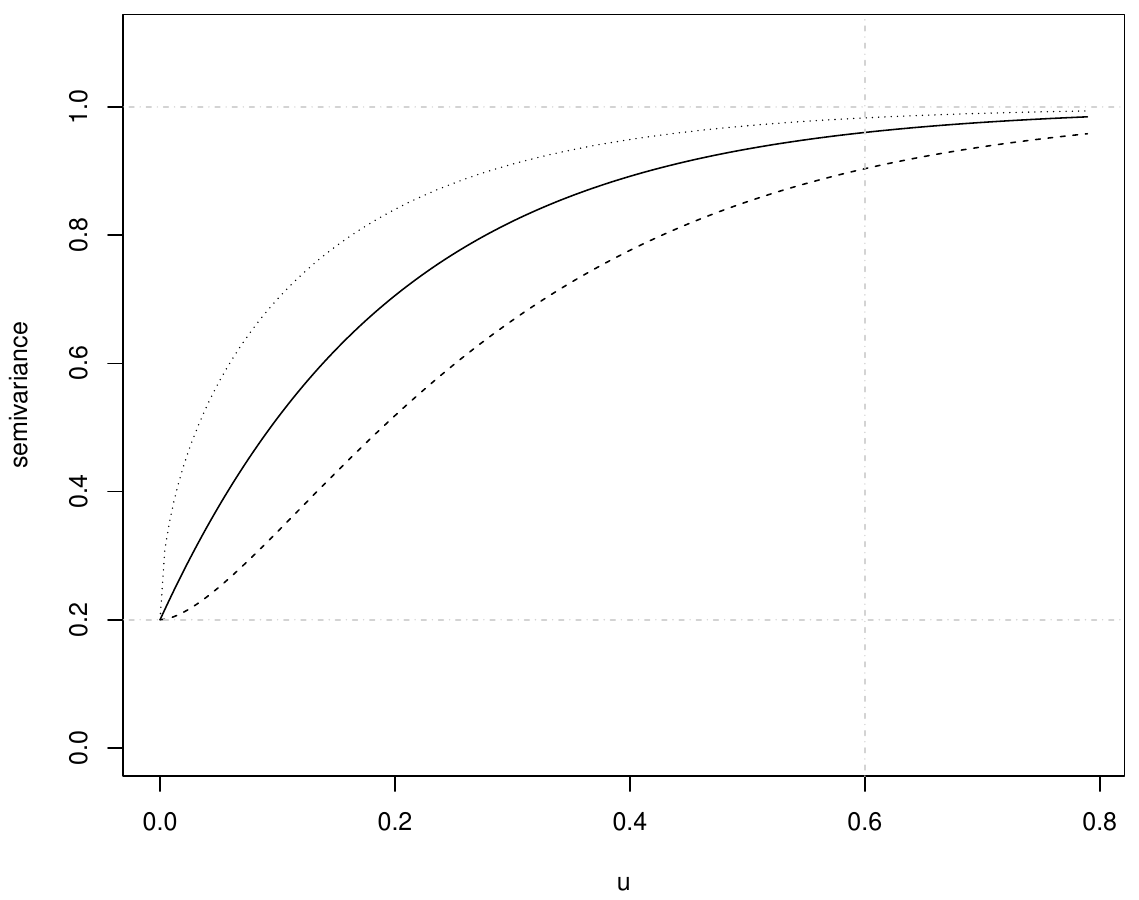}\\ 		
	\end{tabular}
	\caption {(a) Sample and estimation locations (circles and triangles, respectively) for $n_1 = 20\times 20$, (b) theoretical nonlinear trend $\mu_{1}(x_{1},x_{2})$, (c) theoretical nonlinear variance $\sigma_{1}^{2}(x_{1},x_{2})$, and (d) semivariogram models for $c_{0}= 0.2$, $a = 0.6$ and $\nu = 1$ (dashed), $\nu = 0.5$ (solid), $\nu = 0.25$ (dotted line).}   
	\label{fig_nonlin}
\end{figure}

In this section, the heteroscedastic conditional bootstrap procedure described in the previous section is numerically analyzed considering different scenarios. The R \citep{Rsoft} package \texttt{npsp} \citep{npsp} was employed to carry out the simulation experiments. In each case, $N = 1,000$ samples following the model (\ref{hetermodel}) were generated on regular grids in the unit square of sizes $n_1 = 15 \times 15$, $20 \times 20$ and $30 \times 30$. The top right diagonal sites were set as the estimation locations $\mathbf{x}_{\alpha}^e$, $\alpha=1,\ldots,n_0$,  and the remaining ones as the observation sample $\mathbf{x}_i$, $i=1,\ldots,n$ (note that $n = n_1 - n_0$). For example, Figure \ref{fig_nonlin}(a) shows the estimation (triangles) and observation (circles) locations for the case of $n_1 = 20 \times 20$. 

In order to take into account the effect of the functional form of the components of the model (\ref{hetermodel}), the following theoretical trend and variance functions were considered: $\mu_{1}(x_{1},x_{2})=2.5 + \sin(2\pi x_{1})+4(x_{2}-0.5)^{2}$ (nonlinear trend; see Figure \ref{fig_nonlin}(b)),	$\mu_{2}(x_{1},x_{2})=5.8(x_{1} - x_{2} + x_{2}^{2} )$ (polynomial trend), $\mu_{3}(x_{1},x_{2})= 2$ (constant trend), $\sigma_{1}^{2}(x_{1},x_{2}) = (\frac{15}{16} )^2 [1-(2x_1-1)^2]^2 [1-(2x_1-1)^2]^2 + 0.1$ (nonlinear variance; see Figure \ref{fig_nonlin}(c)), $\sigma_{2}^{2}(x_{1},x_{2})= 0.5 (1+ x_{1} + x_{2})$ (linear variance) and $\sigma_{3}^{2}(x_{1},x_{2})=1$ (constant variance, i.e. homocedastic case).
The random errors $\varepsilon(\mathbf{x})$ were generated through a multivariate normal distribution with zero mean, unit variance and an isotropic Mat\'ern variogram model given by:
\begin{equation} \label{variog_matern}
\gamma(\mathbf{u})=
c_{0} + (1 - c_{0}) \left[1-\frac{1}{2^{\nu-1} \Gamma(\nu)} \left( 3\frac{ ||u||}{a} \right)^{\nu}\mathcal{K}_\nu \left( 3\frac{ ||u||}{a} \right) \right],
\end{equation}
where $c_{0}$ denotes the nugget ($1 - c_{0}$ is the partial sill), $a$ is a scale parameter (proportional to the practical
range), and $\mathcal{K}_\nu$ is the second kind modified Bessel function of order $\nu$, being $\nu$ a smoothness parameter. In order to analyze the effect of the spatial dependence, the following parameters have been considered: $c_{0}= 0, 0.2, 0.4, 0.8$, $a = 0.3, 0.6, 0.9$ and $\nu = 0.25, 0.5, 1$. 
Parameter $\nu$ determines the shape of the semivariogram at small lags. For instance, $\gamma(\cdot)$ corresponds to an exponential model when $\nu = 0.5$ (being $a$ its practical range). Figure \ref{fig_nonlin}(d) shows the theoretical semivariograms corresponding to $c_{0}= 0.2$, $a = 0.6$ and the different values $\nu$ considered.


To apply the conditional bootstrap algorithm described in Section \ref{sec_cond_boot}, firstly, it is necessary to estimate the trend $\mu(\cdot)$. For this, we employed the local linear estimator (\ref{trend_est}), with a multiplicative triweight kernel. 
To avoid the bandwidth selection effect in the results, the bandwidth $\mathbf{H}_{{\rm MASE}}$ minimizing the mean average squared error,
$${\rm MASE}(\mathbf{H})=\frac{1}{n}(\mathbf{S}_\mathbf{H}\boldsymbol{\mu}-\boldsymbol{\mu})^t (\mathbf{S}_\mathbf{H}\boldsymbol{\mu}-\boldsymbol{\mu})+\frac{1}{n}\tr(\mathbf{S}_\mathbf{H} \boldsymbol{\Sigma}\mathbf{S}_\mathbf{H}^t),$$
where $\boldsymbol{\mu}=\left[\mu(\mathbf{x_1}), \ldots , \mu(\mathbf{x_n})\right]^t$, was employed for trend estimation. An analogous approach was applied to select $\mathbf{H}_2$ for the variance estimation. 
This also considerably reduced the computing time, as the smoothing matrices $\mathbf{S}_\mathbf{H}$ and $\mathbf{S}_{\mathbf{H}_2}$ only needed to be computed once.
Note that these optimal bandwidths cannot be used in practice, since their calculations depend on the unknown trend $\mu(\cdot)$ and covariance matrix $\boldsymbol{\Sigma}$. In that case, we recommend the use of the CGCV criterion (described in Section \ref{sec_np_modeling}) that provided very similar results to those obtained with $\mathbf{H}_{\rm MASE}$ in some simulation experiments (but it required a much longer computation time because, in addition to the selection of the bandwidths, the corresponding smoothing matrices must be computed in each iteration). 

Once that the trend estimate  $\hat{\mu}_{\mathbf{H}_{{\rm MASE}}}(\mathbf{x})$ is obtained, the iterative algorithm described in Section \ref{sec_np_modeling} is employed to obtain the final variance estimates $\hat{{\sigma}}^2(\cdot)$, the residual semivariogram $\bar{\gamma}_{\hat{\boldsymbol{ \varepsilon}}}(\cdot)$ and its bias-corrected version $\bar{\gamma}(\cdot)$. 
%
%
%

Next, at each simulation the algorithm proposed in Section \ref{sec_boot} was applied with $B=1,000$ bootstrap replicas. For each $\alpha=1,\ldots,n_0$, the conditional probabilities ${r}_c(\mathbf{x}_{\alpha}^e, \mathbf{Y})$ were estimated by $\hat{r}_c(\mathbf{x}_{\alpha}^e, \mathbf{Y})$, given in (\ref{cond_prob_est}), considering threshold values $c=2$, $3$ and $4$. 
At each estimation location $\mathbf{x}_{\alpha}^e$, $\alpha=1,\ldots,n_0$, the squared errors $\left[\hat{r}_c(\mathbf{x}_{\alpha}^e, \mathbf{Y}) - {r}_c(\mathbf{x}_{\alpha}^e, \mathbf{Y})\right]^2$ were computed to evaluate the performance of the proposed procedure.

Note that, taking into account that the responses are normally distributed, the theoretical probabilities ${r}_c(\mathbf{x}_{\alpha}^e, \mathbf{Y})$ can be obtained as:
$$ 1 - \Phi \left[ \frac{ c - \hat{Y}_{SK}(\mathbf{x}_{\alpha}^e)}{\hat{\sigma}_{SK}(\mathbf{x}_{\alpha}^e)} \right],$$
being $\Phi$ the standard normal cumulative distribution function, $\hat{Y}_{SK}(\mathbf{x}_{\alpha}^e)$  the simple kriging prediction of $Y(\mathbf{x}_{\alpha}^e)$, obtained using the theoretical trend and covariance matrix, and $\hat{\sigma}_{SK}^2(\mathbf{x}_{\alpha}^e)$ the corresponding simple kriging variance.

For the sake of brevity, only some representative results are shown here. For example, Table \ref{t1} shows the mean, median and standard deviations of the squared errors ($\times 10^{-2}$) of the estimates obtained with the {proposed} bootstrap approach, for $\mu_{1}$ (nonlinear), $\sigma_{1}^{2}$ (nonlinear), $c_{0}=0.2$, $a=0.6$, $\nu = 0.5$, and the different threshold values and sample sizes considered. It can be observed that, for the different values of $c$, the mean squared error (MSE) decreases as the sample size $n$ increases, suggesting  the consistency of the conditional probability estimator. 

\begin{table}[h!]
 \caption{Mean, median and standard deviations of the squared errors ($\times 10^{-2}$) of the conditional probability estimates, for $\mu_{1}$ (nonlinear), $\sigma_{1}^{2}$ (nonlinear), $c_{0}=0.2$, $a=0.6$, and $\nu = 0.5$}
 \label{t1}
 \begin{center}
	\begin{tabular}{l|ccc|ccc|ccc}
		\hline
		& \multicolumn{3}{|c}{$n_1=15 \times 15$} & \multicolumn{3}{|c}{$n_1=20 \times 20$} & \multicolumn{3}{|c}{$n_1=30 \times 30$} \\ \hline
		$c$ & mean & median & sd & mean & median & sd & mean & median & sd  \\ \hline
		2 & 0.35 & 0.06 & 0.74 & 0.29 & 0.05 & 0.62 & 0.21 & 0.04 & 0.45\\
		3 & 0.66 & 0.03 & 3.58 & 0.46 & 0.02 & 2.44 & 0.28 & 0.01 & 1.36\\
		4 & 0.11 & 0.00 & 0.73 & 0.08 & 0.00 & 0.73 & 0.05 & 0.00 & 0.39\\ \hline					
		\end{tabular}
	\end{center}
\end{table}
	
\begin{figure}[h]
	    \centering
	    \begin{tabular}{cc}
			\footnotesize{(a)} & \footnotesize{(b)} \\
			\includegraphics[scale=0.30]{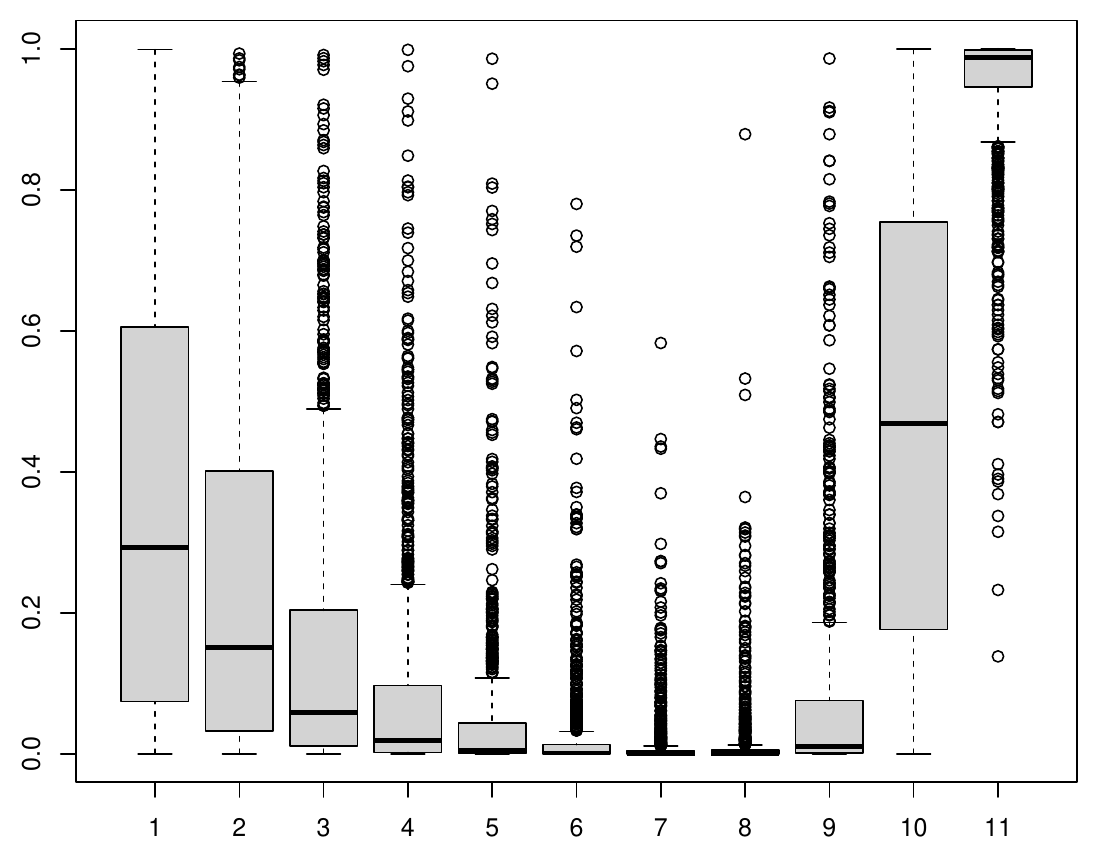}&
			\includegraphics[scale=0.30]{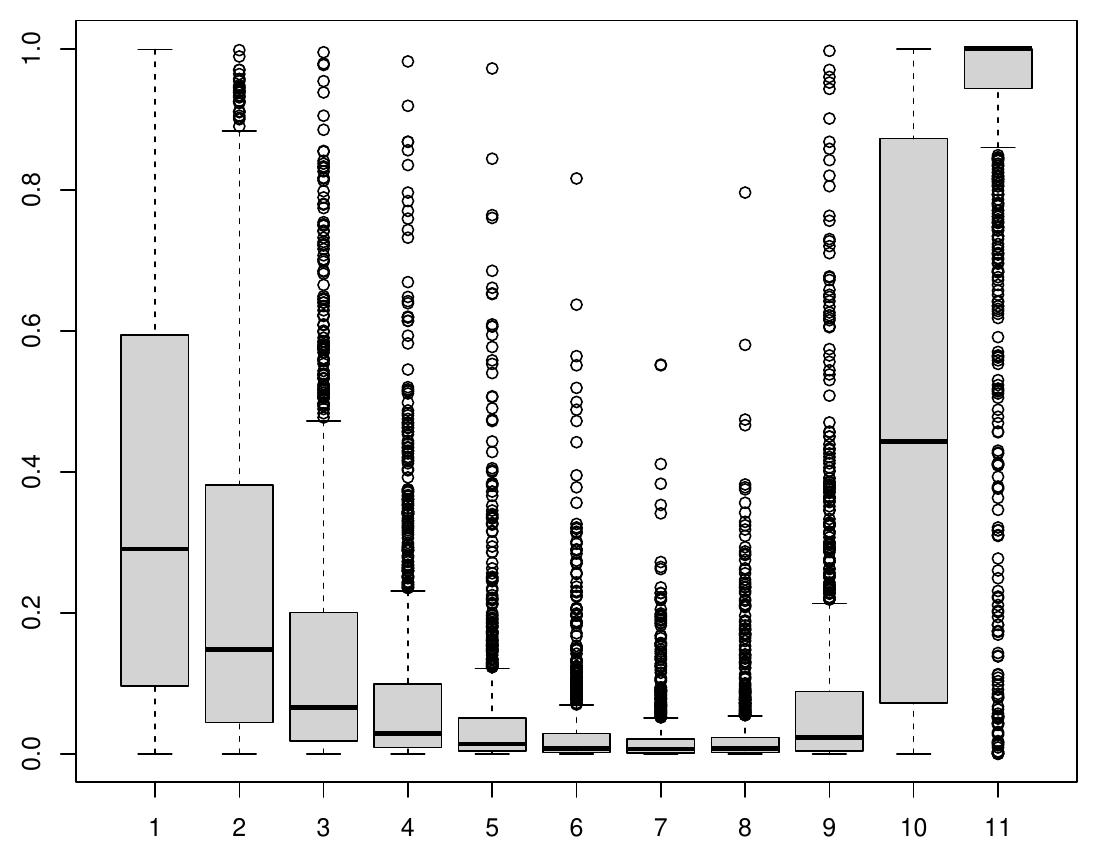}\\ 
		\end{tabular}
	    \caption{Boxplots of the theoretical (a) and estimated (b) conditional probabilities of exceeding a threshold of $c=3$ using the bootstrap method, for $\mu_{1}$ (nonlinear), $\sigma_{1}^{2}$ (nonlinear), $n_1 = 20 \times 20$, $c_{0}=0.2$, $a=0.6$ and $\nu = 0.50$, at the different estimation locations {$\mathbf{x}_{\alpha}^e$, $\alpha=1,\ldots,11$}.}
	    \label{fig_boxplot}
	\end{figure}

The good performance of the proposed approach can also be observed in Figure \ref{fig_boxplot}. It contains boxplots of the theoretical (left panel) and estimated (right panel) conditional probabilities of exceeding a threshold of $c=3$ at the estimation locations $\mathbf{x}_{\alpha}^e$, using the proposed method and considering $\mu_{1}$ (nonlinear), $\sigma_{1}^{2}$ (nonlinear), $n_1 = 20 \times 20$, $c_{0}=0.2$, $a=0.6$ and $\nu = 0.5$. A very similar pattern of the co\-rres\-pon\-ding boxplots for the theoretical and the estimated conditional risks at all estimation locations is observed in this figure. 

The effect of the spatial dependence was also studied by comparing the results obtained with the different values for $a$ and $c_{0}$. In general, an interaction in the effect of these parameters was observed. For example, Table \ref{t2} shows that, for a given level of nugget effect, the error means decrease as the practical range increases. 
As expected, this effect decreases when the nugget is larger (lower spatial dependence), resulting in  similar errors with the different range values.

\begin{table}[h!]
	\caption{Mean, median and standard deviations of the squared errors ($\times 10^{-2}$) of the conditional probability estimates, for $\mu_{1}$ (nonlinear), $\sigma_{1}^{2}$ (nonlinear), $n_1 = 20 \times 20$, $c=3$ and $\nu = 0.5$}
	\label{t2}
	\begin{center}
		\begin{tabular}{l|ccc|ccc|ccc}
	\hline
	& \multicolumn{3}{|c}{$a = 0.3$} & \multicolumn{3}{|c}{$a=0.6$} & \multicolumn{3}{|c}{$a=0.9$} \\ \hline		
	$c_{0}$ & mean & median & sd & mean & median & sd & mean & median & sd  \\ \hline
	0   & 0.57 & 0.03 & 2.39 & 0.51 & 0.01 & 2.60 & 0.40 & 0.00 & 2.34\\
	0.2 & 0.48 & 0.03 & 2.29 & 0.46 & 0.02 & 2.44 & 0.41 & 0.01 & 2.30\\
	0.4 & 0.43 & 0.03 & 2.40 & 0.43 & 0.02 & 2.40 & 0.42 & 0.02 & 2.36\\ 
	0.8 & 0.44 & 0.04 & 2.44 & 0.43 & 0.04 & 2.47 & 0.44 & 0.04 & 2.48\\ 
	\hline					
		\end{tabular}
	\end{center}
\end{table}

Additionally, in the case of a stationary process considering $\mu_{3}(\cdot)$ and $\sigma_{3}^{2}(\cdot)$,
the IK method, briefly mentioned in the Introduction, was also used to estimate the conditional exceeding probabilities 
{and compared with the method proposed in this paper}.
In the IK method only the observed values of the indicator variable $I_{\left\{ Y(\mathbf{x}_{0})\geq c\right\} }$ are considered and, assuming that they are stationary, ordinary kriging is performed to compute predictions at the estimation locations, since $$P\left[ Y(\mathbf{x}_{0})\geq c\mid \mathbf{Y} \right] = \mathbb{E}\left[ I_{\left\{ Y(\mathbf{x}_{0})\geq c\right\} }\mid \mathbf{Y} \right],$$ 
\citep[see, e.g.][Section 6.3.3, for further details]{C12}. In practice, this parametric approach was implemented using the \texttt{geoR} package \citep{geor}, fitting an exponential variogram model to the indicator variable (assuming a constant trend).
The performance of this method was compared with that obtained by the proposed approach in this scenario.
Note that the nonparametric procedure was applied assuming (wrongly) the presence of non-constant trend and variance functions.

As an illustrative example of the comparison {results}, Table \ref{t3a} shows a summary of the squared errors ($\times 10^{-2}$) of the estimates obtained with the IK and the proposed approach (denoted by NP 
{in this table}), for $n_1 = 20 \times 20$, $a = 0.6$, $c_0 = 0.2$ and different values of $\nu$. 
In general, the errors obtained by the proposed method are lower than those provided by the IK 
{approach}. This 
{could be because of} another limitation of 
{the IK} method, due to the loss of information by using only the discretized response values \citep[see, e.g.][]{T08}. Therefore, even when a non-constant trend and variance are incorrectly assumed, the nonparametric proposed method seems to be a better alternative for estimating the conditional probability.

\begin{table}[h!]
	\caption{Mean, median and standard deviations of the squared errors ($\times 10^{-2}$) obtained with the IK and the proposed (denoted by NP) methods, for $\mu_{3}(\cdot)=2$ (constant), $\sigma_{3}^{2}(\cdot)=1$ (homoscedastic), $n_1 = 20 \times 20$, $c_{0}=0.2$, $a=0.6$ and the different $\nu$ values. }
	\label{t3a}
	\begin{center}
		\begin{tabular}{|c|c|ccc|ccc|}
			\hline
			\multicolumn{2}{|c}{Method} & \multicolumn{3}{|c}{IK} & \multicolumn{3}{|c|}{NP} \\ 			\hline		
			$\nu$ & $c$ & mean & median & sd & mean & median & sd \\ \hline
				    &2	&0.88	&0.39	&1.29	&0.22	&0.07	&0.42 \\
			0.25	&3	&0.60	&0.20	&1.13	&0.12	&0.03	&0.34 \\
					&4	&0.13	&0.01	&0.58	&0.02	&0.00	&0.15 \\ \hline
					&2	&1.07	&0.42	&1.64	&0.23	&0.06	&0.49 \\
			0.5		&3	&0.73	&0.14	&1.55	&0.13	&0.01	&0.40 \\
					&4	&0.16	&0.00	&0.81	&0.03	&0.00	&0.23 \\ \hline
					&2	&0.85	&0.25	&1.52	&0.17	&0.03	&0.49 \\
			1		&3	&0.56	&0.05	&1.44	&0.10	&0.01	&0.36 \\
					&4	&0.12	&0.00	&0.70	&0.03	&0.00	&0.29 \\ \hline
		\end{tabular}
	\end{center}
\end{table}

As a finally study, the case of irregular sampling was analyzed. The previous scenarios were considered, but now $n_1$ spatial locations were randomly generated following a bidimensional uniform distribution over the unit square. The same $n_0$ estimation locations as in the regular sampling design were chosen. 
 In general, very similar results 
 {to those achieved under regular design} {were} obtained, although the errors {were} slightly smaller with the irregular design. 
 For instance, Table \ref{t4} shows the MSE for $n_1 = 20 \times 20$, $c_{0}=0.2$, $a=0.6$, $c=3$ and the different values of the smoothness parameter $\nu$.
 As it can be seen, by increasing the smoothness of the process $\nu$ the errors tend to be smaller (similar results are observed in Table \ref{t3a}).
 Additionally, errors tend to be larger when more complex models are considered.

\begin{table}[h!]
	\caption{Averaged squared errors ($\times 10^{-2}$) of the conditional probability estimates under irregular sampling, for $n_1 = 20 \times 20$, $c_{0}=0.2$, $a=0.6$ and $c=3$.}
	\label{t4}
	\begin{center}
		\begin{tabular}{l|ccc|ccc|ccc}
			\hline
			& \multicolumn{3}{|c}{$\sigma_{1}^{2}$ (nonlinear)} & \multicolumn{3}{|c}{$\sigma_{2}^{2}$ (linear)} & \multicolumn{3}{|c}{$\sigma_{3}^{2}$ (constant)} \\ \hline		
			$\nu$ (Matèrn Model) & 0.25 & 0.50 & 1.00 & 0.25 & 0.50 & 1.00 & 0.25 & 0.50 & 1.00  \\ \hline
			$\mu_{1}$ (nonlinear) &0.50	&0.46	&0.39	&0.21	&0.18	&0.14	&0.23	&0.20	&0.16\\
			$\mu_{2}$ (polynomial) &0.16	&0.14	&0.10	&0.09	&0.09	&0.09	&0.07	&0.07	&0.08\\
			$\mu_{3}$ (constant) &0.12	&0.13	&0.11	&0.13	&0.15	&0.16	&0.12	&0.13	&0.15\\
			\hline					
		\end{tabular}
	\end{center}
\end{table}

\section{Application to precipitation data}
\label{sec_app_data}

\begin{figure}[ht]
    \centering
    	\begin{tabular}{cc}
			\footnotesize{(a)} & \footnotesize{(b)} \\
			{\includegraphics[height=3cm]{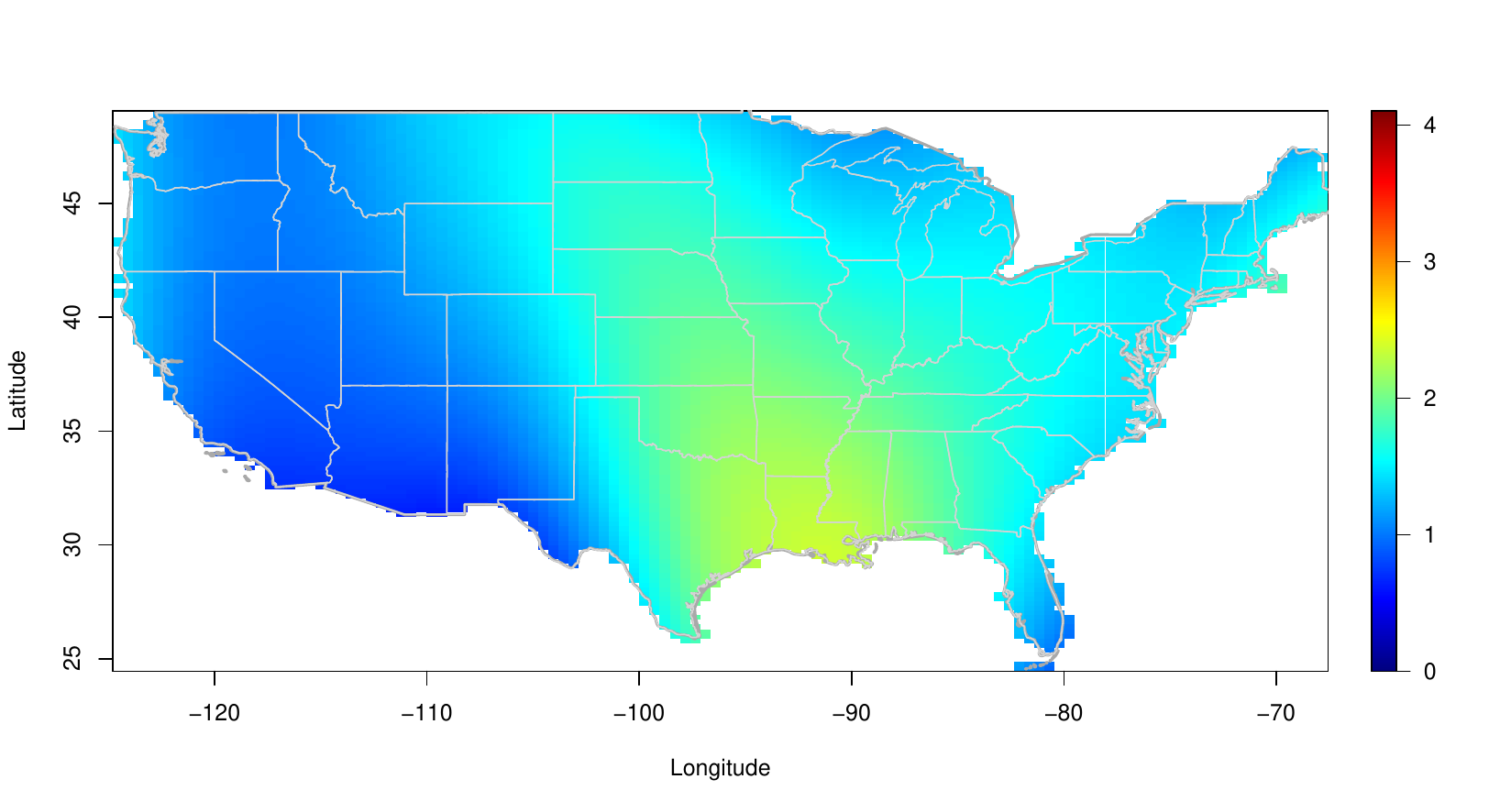}} & {\includegraphics[height=3cm]{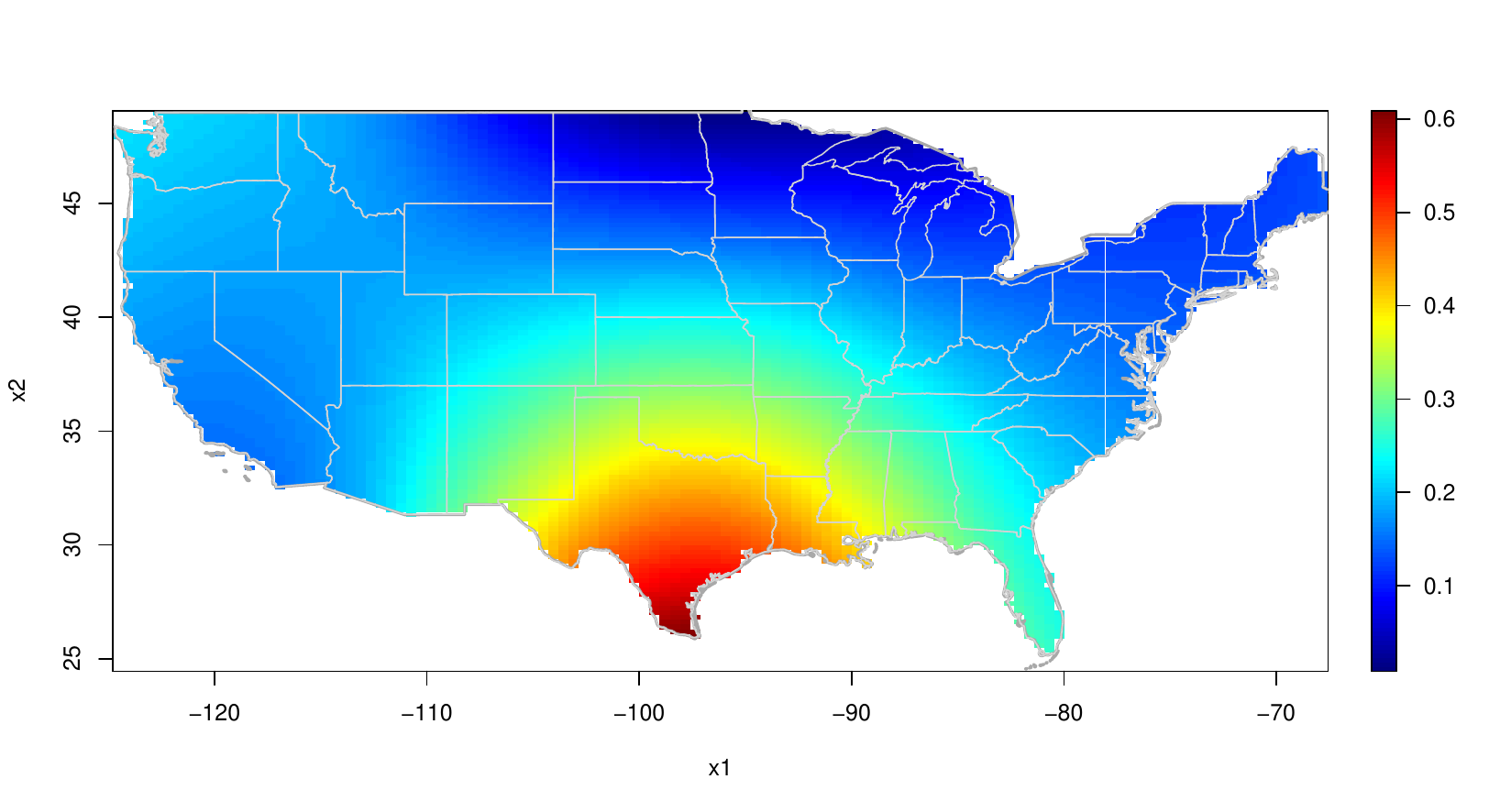}} \\
			\multicolumn{2}{c}{\footnotesize{(c)}}  \\
			\multicolumn{2}{c}{\includegraphics[height=3cm]{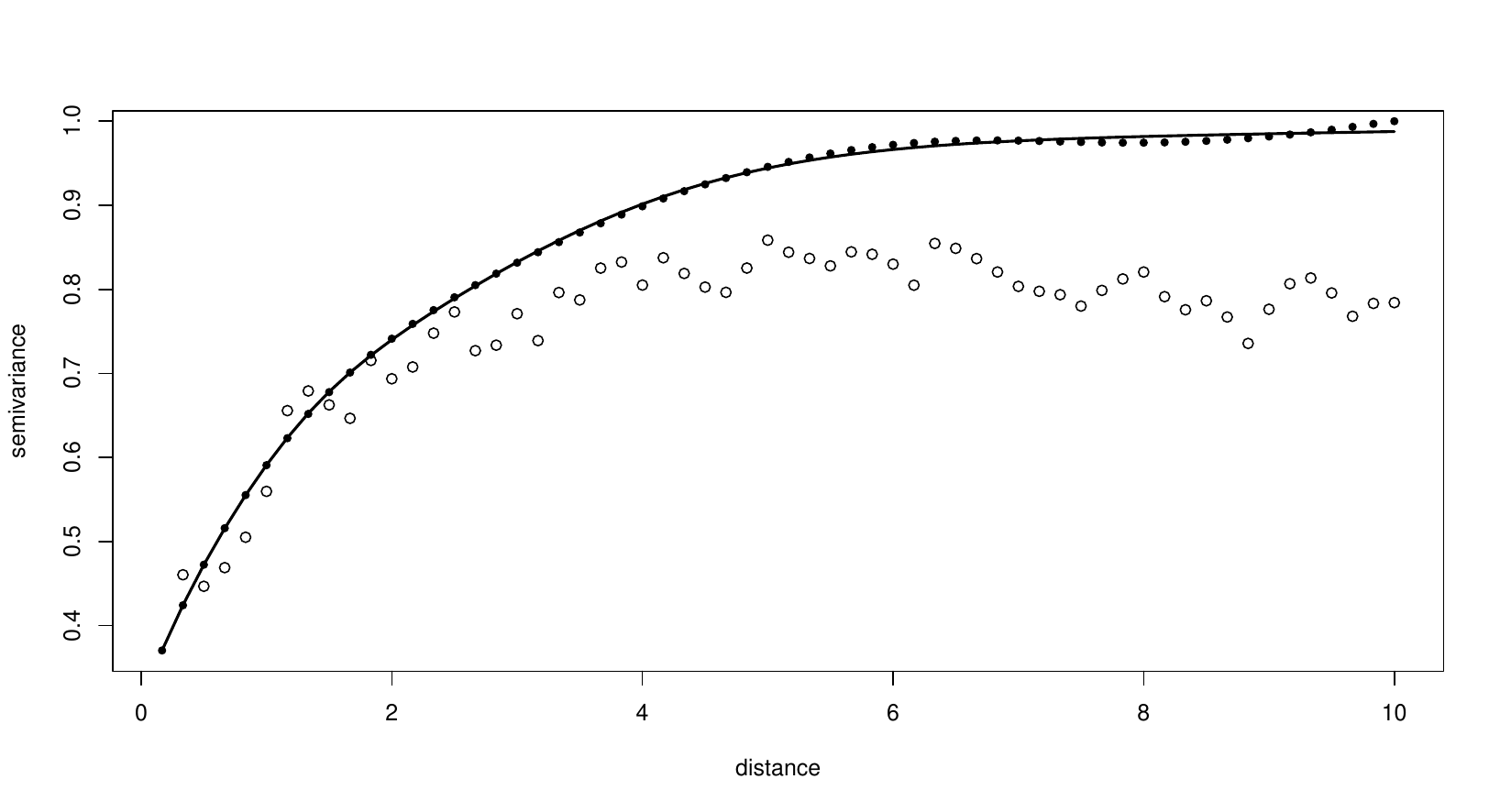}} \\ 
		\end{tabular}
    \caption{Estimated nonparametric trend (a), conditional variance (b) and semivariogram (c) of the precipitation data (in root-squared rainfall inches).}
    \label{fig_rain_np_estim}
\end{figure}

In order to illustrate the performance in practice of the proposed methodology, the precipitation data set briefly mentioned in the Introduction is considered. This data set is supplied with the \texttt{R} package \texttt{npsp}. 
The trend and variogram estimates were obtained using the iterative algorithm described in Section \ref{sec_np_modeling}.
The final trend and variance function estimates are shown in Figure \ref{fig_rain_np_estim}(a) and (b) respectively. The Figure \ref{fig_rain_np_estim}(c) shows the pilot residual variogram $\hat{\gamma}_{\hat{\boldsymbol{ \varepsilon}}}(\cdot)$ (circles) and the bias-corrected estimate $\bar{\gamma}(\cdot)$ (solid line).

\begin{figure}[ht]
    \centering
    \begin{tabular}{cc}
		\footnotesize{(a)} & \footnotesize{(b)} \\
		\includegraphics[height=3cm]{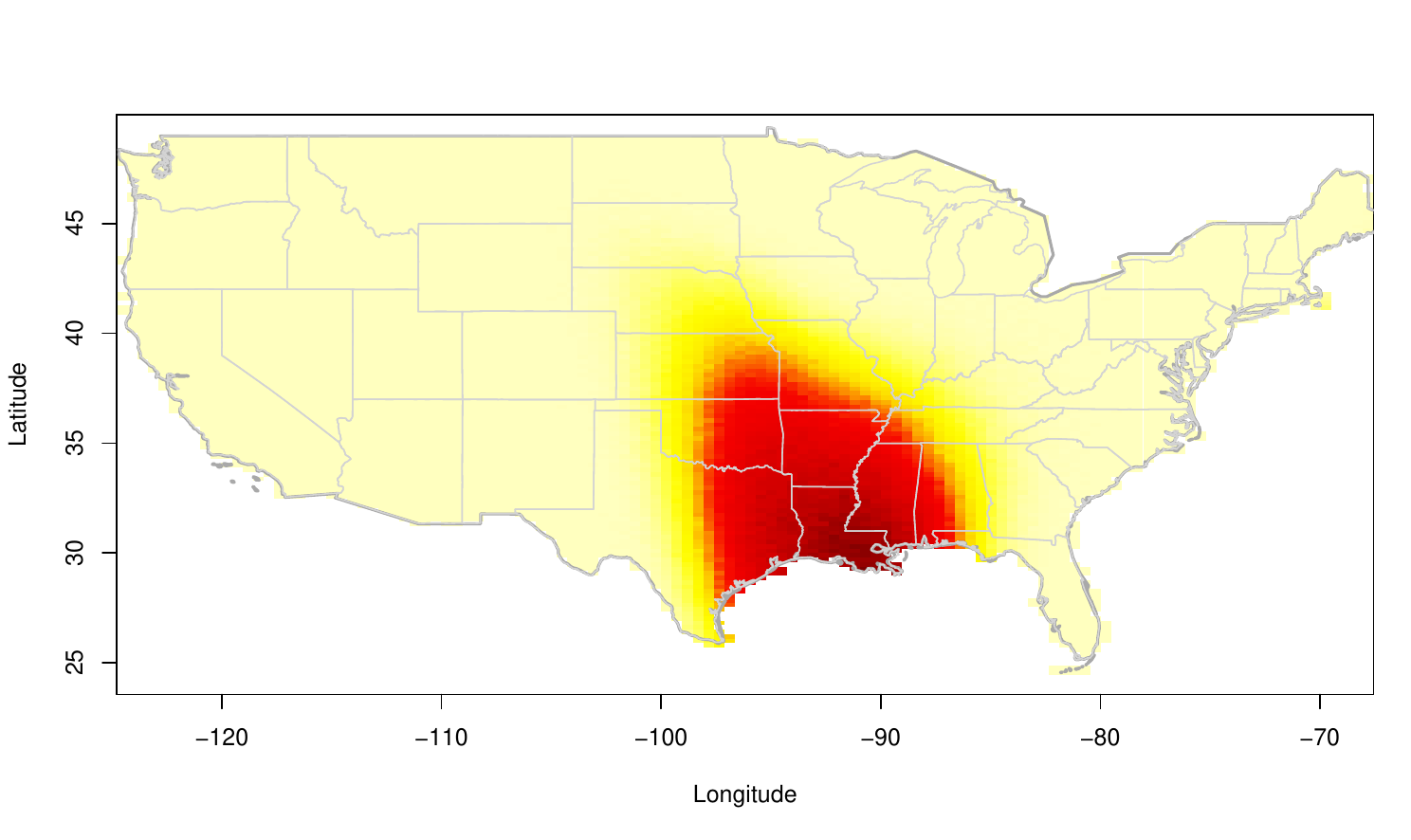}&
		\includegraphics[height=3cm]{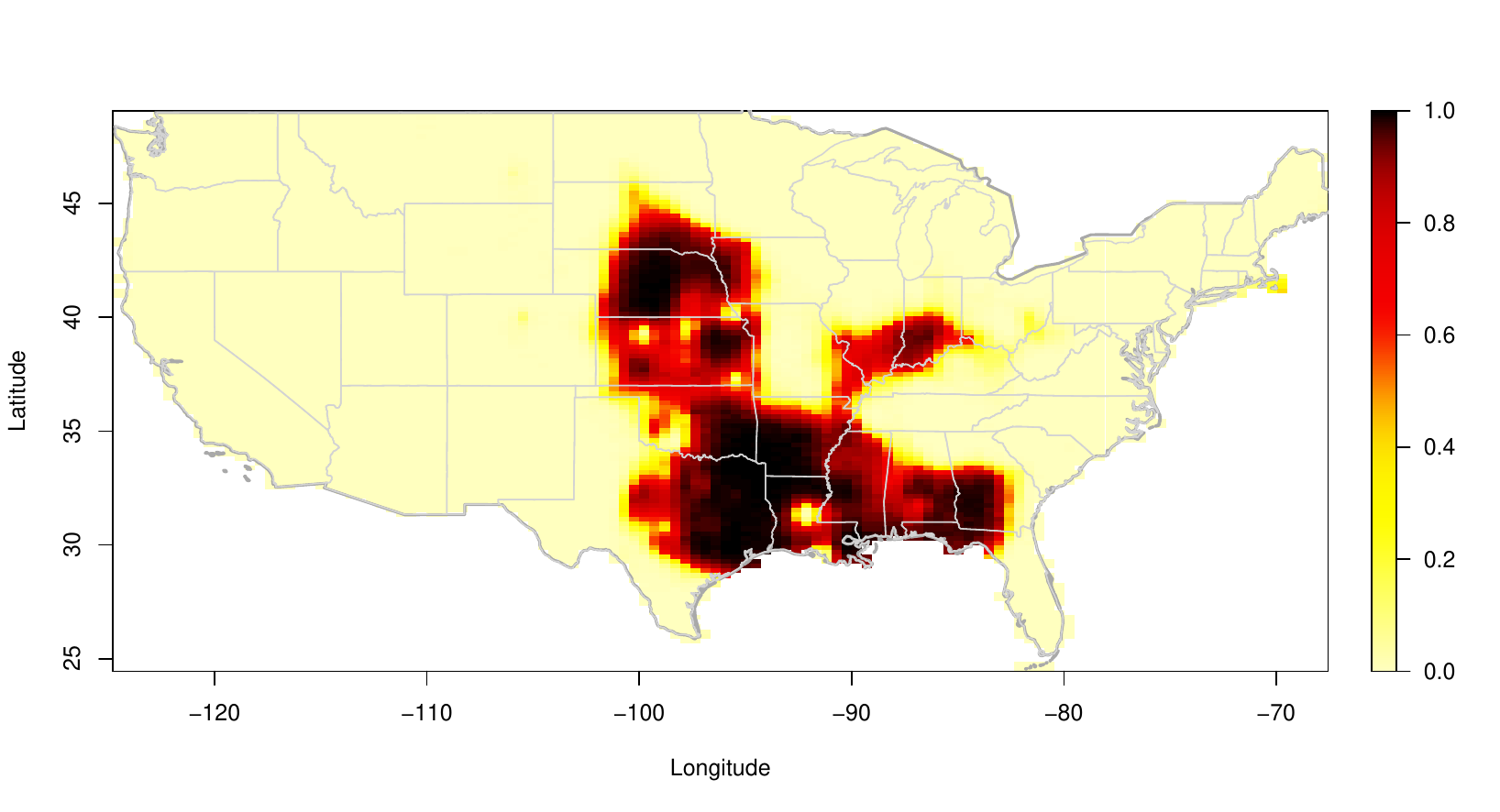}\\ 
	\end{tabular}
    \caption{Estimated unconditional (a) and conditional (b) risk maps for $c=2.0$.}
    \label{fig_risk_rain}
\end{figure}

Estimated probability maps were computed for different threshold values (square-root of rainfall inches), $c = \{1.0, 1.5, 2.0, 2.5, 3.0, 3.5, 4.0\}$, by applying the two bootstrap algorithms (unconditional and conditional) described in Section \ref{sec_uncond_boot} and \ref{sec_cond_boot} respectively, with $B=1000$ bootstrap replicas in each case. For reason of space, only the case of threshold $c=2.0$ is included here (Figure \ref{fig_risk_rain}).

In Figure \ref{fig_risk_rain}, it can be observed that the unconditional bootstrap provides smoother estimates than those obtained with the conditional approach, emphasizing the dominant effect of the trend estimate (large-scale variability). On the other hand, the conditional map shows higher variability, reaching the extreme values 0 and 1 at sample locations. These differences are more evident in the northern regions of the central area of the map, where the proposed method produces estimates in line with the observed values (shown in Figure \ref{fig_obser}), due to the stronger effect of the spatial dependence (small-scale variability) on the conditional estimates.

\section{Discussion and further remarks}
\label{sec_conclus}

A bootstrap algorithm to estimate threshold exceeding conditional probabilities under heteroscedasticity of the spatial process is proposed and numerically analyzed in this paper. 
The probabilities are approximated from bootstrap conditional replicates obtained in a two-stage procedure. In the first step, the unconditional bootstrap method proposed in \cite{FC20} is used.
In the second step, the unconditional replicates are combined with kriging predictions so that the resulting values coincide with the observed values at the sample locations. 

Unlike traditional methods, such as IK or DK, the new approach is designed to be applied for processes that are not stationary (in the mean or in the variance). Moreover, the new approach is fully nonparametric and, therefore, problems due to model misspecification are at least partially avoided. 
However, to ensure the consistency of the local linear estimator, smooth functions for the trend, variance and variogram are being implicitly  assumed.
Additionally, a ``bias-corrected'' method to jointly estimate the variance function and the spatial dependence is employed.
In this way, the proposed bootstrap algorithm takes into account that the variability of the residuals is not equal to that of the true errors.
Note that although the local linear estimator has been considered in this research due to its good properties, other linear smoothers could also be used.

The complete simulation study shows a good behavior of the new method and its appropriate performance in different scenarios, considering several degrees of spatial dependence and functional forms for the spatial trend and variance.
Simulations considering regular and non-regular designs were performed. The results obtained in both frameworks were very similar and analogous conclusions could be deduced from them. For this reason, for the sake of brevity, only some representative scenarios in the case of irregular designs are included in the paper.
Note that the case of non-regular design is also illustrated in the real data application in Section \ref{sec_app_data}.
It is important to remark that in the case of non-regular designs the computational cost of the simulations is much larger, as the optimal bandwidths and the corresponding smoothing matrices, for trend and variance estimation, have to be computed in each iteration (see the comments about bandwidth selection in Section \ref{sec_simulation}).
		
The numerical analysis carried out in this research was performed with the statistical environment \texttt{R} \citep{Rsoft}, using the functions for nonparametric regression and variogram estimation supplied with the \texttt{npsp} package \citep{npsp}.

 \section*{Acknowledgements}\label{acknowledgements}
The research of Rub\'{e}n Fern\'{a}ndez-Casal and Mario Francisco-Fer\-n\'{a}n\-dez has been supported by grant PID2020-113578RB-I00, funded by MCIN/AEI/10.13039/501100011033/. It has also been supported by the Xunta de Galicia (Grupos de Referencia Competitiva ED431C-2020/14) and by CITIC that is supported by Xunta de Galicia, convenio de colaboraci\'on entre la Conseller\'ia de Cultura, Educaci\'on, Formaci\'on Profesional e Universidades y las universidades gallegas para el refuerzo de los centros de investigaci\'on del Sistema Universitario de Galicia (CIGUS). The research of Sergio Castillo P\'aez has been supported by the Universidad de las Fuerzas Armadas ESPE, from Ecuador. The authors thank the Editor, an Associate Editor and  two anonymous referees for numerous useful comments that significantly improved this article.

\bibliographystyle{spbasic}

\end{document}